\documentclass[english]{article}
\usepackage[T1]{fontenc}
\usepackage[latin9]{inputenc}
\usepackage{geometry}
\usepackage{color,times}
\usepackage{amsmath,ulem}
\usepackage{amssymb}
\usepackage{stackrel}
\usepackage{graphicx}
\usepackage{esint,epsfig}

\makeatletter

\providecommand{\tabularnewline}{\\}

\date{}
\usepackage{cite}

\makeatother

\usepackage{babel}
\begin{document}
\title{Can we control the amount of useful nonclassicality in a photon added
hypergeometric state?}
\author{{\normalsize{}Priya Malpani$^{\mathparagraph}$, Kishore Thapliyal$^{\ddagger,}$}\thanks{Email: kishore.thapliyal@upol.cz}{\normalsize{},
Anirban Pathak$^{\mathparagraph,}$}\thanks{Email: anirban.pathak@jiit.ac.in}{\normalsize{}
}\\
{\normalsize{} $^{\mathparagraph}$Jaypee Institute of Information
Technology, A-10, Sector-62, Noida, UP-201309, India}\\
{\normalsize{}$^{\ddagger}$RCPTM, Joint Laboratory of Optics of Palacky
University and Institute of Physics}\\
{\normalsize{}of Academy of Science of the Czech Republic, Faculty
of Science, Palacky University, }\\
{\normalsize{} 17. listopadu 12, 771 46 Olomouc, Czech Republic}}
\maketitle
\begin{abstract}
Non-Gaussianity inducing operations are studied in the recent past
from different perspectives. Here, we study the role of photon addition,
a non-Gaussianity inducing operation, in the enhancement of nonclassicality
in a finite dimensional quantum state, namely hypergeometric state
with the help of some quantifiers and measures of nonclassicality.
We observed that measures to characterize the quality of single photon
source and anticlassicality lead to the similar conclusion, i.e., to
obtain the desired quantum features one has to choose all the state
parameters such that average photon numbers remains low. Wigner logarithmic
negativity of the photon added hypergeometric state and concurrence
of the two-mode entangled state generated at the output of a beamsplitter
from this state show that nonclassicality can be enhanced by increasing
the state parameter and photon number addition but decreasing the
dimension of the state. In principle, decreasing the dimension of
the state is analogous to holeburning and is thus expected to increase
nonclassicality. Further, the variation of Wigner function not only
qualitatively illustrates the same features as observed quantitatively
through concurrence potential and Wigner logarithimic negativity,
but illustrate non-Gaussianity of the quantum state as well. 
\end{abstract}

\section{Introduction}

On the verge of second quantum revolution \cite{dowling2003quantum},
nonclassical states are essential elements to establish quantum dominance.
Literature enlightens, a lot of work has been performed on the study
of generation of various nonclassical states and their properties
(\cite{pathak2018classical,tan2019nonclassical,agarwal2012quantum}
and references therein). To be specific, the quantum states with non-positive
Glauber-Sudarshan $P$ function \cite{glauber1963coherent,sudarshan1963equivalence},
called nonclassical states, are at the core of the second quantum
revolution \cite{dowling2003quantum,grimm2005quantum}. The nonclassical
states are extremely important for obtaining quantum advantage in
computing \cite{cacciapuoti2019quantum,lund2008fault,mcmahon2007quantum},
communication \cite{bennett1992communication,bennett1993teleporting},
cryptography \cite{bennett1984quantum}, metrology \cite{giovannetti2006quantum,giovannetti2011advances},
simulation \cite{georgescu2014quantum,marques2015experimental}, sensing
\cite{pirandola2018advances,degen2017quantum}, etc. For instance,
many of these applications require squeezed states, having uncertainty
in one of the quadratures less than the corresponding value for coherent
state; entangled states, which cannot be written as a product of quantum
states of individual subsystems; and on-demand single photon sources,
which are necessarily antibunched states. 

Generally, a finite dimensional quantum state is referred to as qudit
(in a $d$-dimensional Hilbert space), which can be expressed in the
Fock basis as
\begin{equation}
\left|\psi_{d}\right\rangle =N_{d}\sum_{n=0}^{d-1}c_{n}\left|n\right\rangle ,\label{eq:qudit}
\end{equation}
where $N_{d}$ is the normalization constant, and $\left|c_{n}\right|^{2}=p_{n}$
is the photon number distribution of the state. Thus, we know $p_{n}=0\,\forall n\geq d$,
which is termed as a hole in the photon number distribution. Additionally,
we know in terms of the Glauber-Sudarshan $P$ function, we can write
the photon number distribution as 
\begin{equation}
p_{n}=\int P\left(\alpha\right)\left|\left\langle n|\alpha\right\rangle \right|^{2}d^{2}\alpha,\label{eq:pnd}
\end{equation}
which allows us to conclude that $P\left(\alpha\right)<0$ if $p_{n}=0$
for any value of $n$. The motivation to study finite dimensional
(qudit) states lies in the fact that some of these states can be reduced
mathematically to the most nonclassical (Fock state) as well as the
most classical (coherent state) states in the limiting cases. Most
significant examples of finite dimensional (qudit) states worth mentioning
are a class of intermediate states, such as hypergeometric \cite{fu1997hypergeometric},
binomial \cite{stoler1985binomial}, negative binomial \cite{barnett1998negative},
vacuum filtered binomial \cite{malpani2020manipulating}, negative
hypergeometric \cite{fan1998negative}, photon added binomial \cite{malpani2020manipulating},
shadowed \cite{srinivasan1996shadowed} and shadowed-like \cite{lee1997squeezing}
negative binomial states, etc. Initially the investigations on these
states were limited to theoretical studies, but with the advent of
advanced experimental techniques, a few groups have succeeded in generating
such states \cite{franco2006single,franco2010efficient,giordani2019experimental}
and possibly more to follow. Due to several applications of qudit
states various nonclassical properties (both lower- and higher-order
antibunching, squeezing, sub-Poissonian photon statistics, etc.) have
been studied very rigorously \cite{lee1997squeezing,miranowicz2014phase,alam2018higher1,malpani2020manipulating}.
However, most of these works were focused on the witnesses of nonclassicality
and relatively less effort has been given on the quantitative aspects
of nonclassicality. 

Hypergeometric state is a one-parameter generalization of binomial
state and is also studied in reference to deformed oscillator algebra
\cite{fu1997hypergeometric}. Non-Gaussianity inducing operations,
such as photon addition, subtraction, vacuum filtration, quantum scissors,
photon catalysis are used as nonclassicality inducing and enhancing
operations in several quantum state engineering proposals. Specifically,
advantage of photon addition and subtraction in application of quantum
phase in quantum metrology \cite{malpani2019quantum}, quantum scissors
in teleportation and cryptography \cite{leonski2011quantum,goyal2013teleporting,ghalaii2020discrete},
entanglement purification (\cite{hu2019entanglement} and references
therein), nonclassicality and non-Gaussianity enhancement \cite{malpani2019lower,malpani2019quantum,malpani2020manipulating},
are reported recently. Further, non-Gaussian states obtained by such
quantum state engineering operations have applications in continuous
variable quantum cryptography and computation (\cite{walschaers2018tailoring,srikara2020continuous,saxena2020continuous}
and references therein). The continuous variable quantum cryptography
is considered a better candidate than corresponding discrete variable
counterparts for a metropolitan network \cite{pirandola2015reply}
and non-Gaussian operations can further enhance this performance (for
example, see \cite{guo2019continuous,hu2020continuous}).
Moreover, recently engineered qudit states have been used experimentally
in quantum walk \cite{giordani2019experimental}, which may be a useful
resource for quantum walk based quantum cryptographic schemes \cite{vlachou2018quantum}.

Motivated by above mentioned facts, here, we aim to study the role
of photon addition in the enhancement of nonclassicality in hypergeometric
state. The choice of this state is particularly important as it holds
the potential to be used for realizing all the known applications
of finite dimensional states in quantum technology, and a large set
of finite and infinite dimensional quantum states can be reduced in
the limiting cases of the hypergeometric state (discussed in detail
in next section). {  Specifically, in this paper, we aim to study nonclassical
features of photon added hypergeometric state (PAHS), in terms of
a few quantitative measures of quantumness, namely characterization
of the quality of the single photon source, anticlassicality, concurrence
potential, and Wigner logarithmic negativity.} We will
further reduce the corresponding results for a set of other quantum
states, composing hypergeometric, binomial, photon added binomial,
coherent, and photon added coherent states, in the limiting cases.

The rest of the paper is organized as follows. In Section \ref{part:State-to-bePAHS},
we discuss all the properties and Wigner
function of PAHS and its limiting cases.  We summarize
our findings regarding nonclassicality quantifiers to characterize
single photon source, anticlassicality, concurrence potential, and Wigner logarithmic negativity in the subsequent section.
Finally, we conclude our results in Section \ref{sec:Result-and-concluding}.

\section{State to be investigated: Photon added hypergeometric state\label{part:State-to-bePAHS}}

Hypergeometric state was introduced by Fu and Sasaki \cite{fu1997hypergeometric}
in 1996. It can be expressed as the superposition of number states
in the $M+1$-dimensional space analogous to Eq. (\ref{eq:qudit})
as\textcolor{red}{{} }

\begin{equation}
\left|L,M,\eta\right\rangle =\sum_{n=0}^{M}\left[\tbinom{L\eta}{n}\tbinom{L\left(1-\eta\right)}{M-n}\right]^{\frac{1}{2}}\tbinom{L}{M}^{-\frac{1}{2}}\left|n\right\rangle ,\label{HS}
\end{equation}
where probability $\eta\in\left[0,1\right]$ and $L\geq\max\left\{ M\eta^{-1},M\left(1-\eta\right)^{-1}\right\} $
are the real parameters. Also, $\tbinom{N}{l}$ is the binomial coefficient.
The quantum state has photon number distribution $\left|\left\langle n\mid L,M,\eta\right\rangle \right|^{2}=\tbinom{L\eta}{n}\tbinom{L\left(1-\eta\right)}{M-n}\tbinom{L}{M}^{-1}$
defined as hypergeometric distribution.  {Note that the hypergeometric distribution is closely related to the binomial distribution. The former involves the probability of $n$ successful cases in $M$ draws without replacement from a total population of $L$ with only $L\eta$ having the desired feature. While the latter involves the probability of $n$ successful cases in $M$ draws with replacement. Thus, in the limits of $L\rightarrow\infty$, the total population size would not change with each draw and hypergeometric distribution reduces to the binomial distribution.}

This quantum state can be transformed to obtain $k$-PAHS by repeatedly
applying a creation operator on the hypergeometric state. The analytical
expression in this case is given by 

\begin{equation}
\left|L,M,\eta,k\right\rangle =a^{\dagger k}\left|L,M,\eta\right\rangle =N_{{\rm PAHS}}\sum_{n=0}^{M}\left[\tbinom{L\eta}{n}\tbinom{L\left(1-\eta\right)}{M-n}\right]^{\frac{1}{2}}\tbinom{L}{M}^{-\frac{1}{2}}\sqrt{\frac{\left(n+k\right)!}{n!}}\left|n+k\right\rangle .\label{PAHS}
\end{equation}
where $k$ is the number of photons added. We need to renormalize
the state as photon addition is a non-unitary operation, and the normalization
constant for PAHS can be computed as

\begin{equation}
N_{{\rm PAHS}}=\left\{ \sum_{n=0}^{M}\left[\tbinom{L\eta}{n}\tbinom{L\left(1-\eta\right)}{M-n}\right]\tbinom{L}{M}^{-1}\frac{\left(n+k\right)!}{n!}\right\} ^{-\frac{1}{2}}.\label{Norm}
\end{equation}
Lower- and higher-order antibunching, sub-Poissonian photon statistics
and squeezing in hypergeometric state are reported in \cite{fu1997hypergeometric}.
{  Here, our objective is to investigate the role of photon addition
in enhancement in the nonclassical features of the state qualitatively using Wigner function
\cite{wigner1932quantum} as well as with respect
to a set of measures to characterize the quality of single photon source \cite{pathak2010recent},
anticlassicality \cite{dodonov2003classicality}, concurrence potential
\cite{wootters2001entanglement,asboth2005computable}, and Wigner logarithmic negativity \cite{albarelli2018resource}. }

\begin{figure}
\centering{}\includegraphics[scale=0.6]{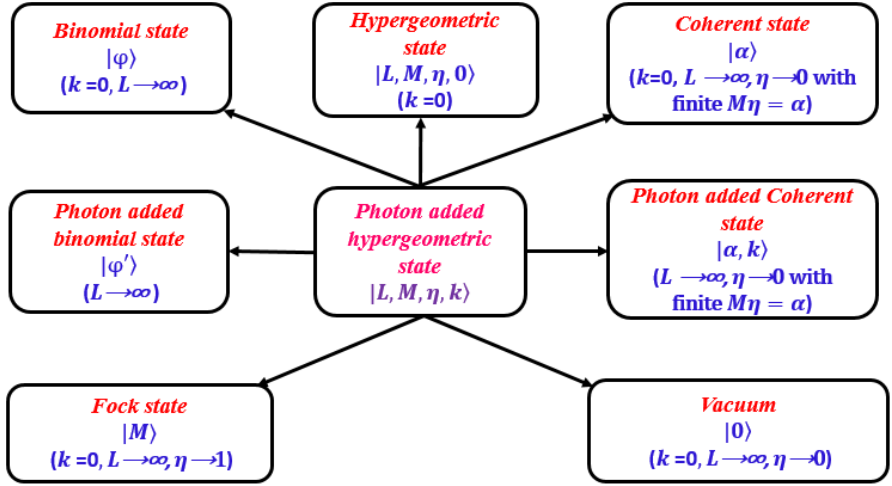}\caption{\label{fig:PASHS}(Color online) The complete set of quantum states
that can be reduced from PAHS in the limiting cases. Here, $\left|\varphi\right\rangle $
and $\left|\varphi'\right\rangle $ corresponds to binomial and photon
added binomial states, respectively.}
\end{figure}

Before we proceed further it is imperative to mention that the PAHS
can be reduced to a set of finite and infinite dimensional quantum
states in limiting cases (as summarized in Fig. \ref{fig:PASHS}).
For instance, hypergeometric state can be obtained from PAHS for $k=0$.
As hypergeometric state can be reduced to the binomial state $\left|\varphi\right\rangle $
in the limits of $L\rightarrow\infty$, so PAHS gives us photon added
binomial state $\left|\varphi'\right\rangle $ in that limit. Further,
considering $M\eta=\alpha$ in the limits of $M\rightarrow\infty$,
we can also obtain coherent state and photon added coherent state
with real $\alpha$ parameter from binomial and photon added binomial
states, respectively. Interestingly, Fock state $\left|M\right\rangle $
and vacuum $\left|0\right\rangle $ can also be obtained from the
binomial state by considering $\eta=1$ and $\eta=0$, respectively.
These limiting cases add value to the present study by clearly indicating
a general nature of the present study. Specifically, as special cases
of the present study, nonclassical and non-Gaussian features of the
limiting states can be visualized and quantified with ease. Some of
these states are already shown to be useful in continuous variable
quantum computation and communication \cite{grosshans2002continuous,pinheiro2013quantum},
quantum phase estimation \cite{izumi2016optical}, implementation
of CNOT gate \cite{franco2010efficient}, etc.

Generation of different types of binomial states is proposed theoretically \cite{moussa1998generation} as well as performed experimentally \cite{franco2006single,franco2010efficient,giordani2019experimental}. Specifically, an intermediate state can be prepared in a cavity using Jaynes-Cummings Hamiltonian by passing $M$ atoms through the cavity initially at vacuum (\cite{rev} and references therein). The experimental parameters can be obtained numerically to generate target state at a target time with the post-selection of all the atoms in the ground state (which will disentangle it from the cavity mode) \cite{gen}. Therefore, using the same method the hypergeometric state can be prepared. Further, $k$-photon addition on the cavity mode in hypergeometric state can be performed by sending $k$ atoms prepared in the excited state through the cavity and measured subsequently in the ground state \cite{PACS}.  Independently, photon addition can be performed using spontaneous parametric down conversion and conditioning on the measurement of one of the down converted modes \cite{PACS-ex}.

\subsection{Wigner function of the state \label{subsec:Wigner-function}}

Nonclassicality in phase space is represented by the negative values
of quasiprobability distributions \cite{agarwal2012quantum,scully1999quantum},
such as Glauber-Sudarshan $P$ function \cite{glauber1963coherent,sudarshan1963equivalence},
Wigner function \cite{wigner1932quantum}. Here, we discuss the Wigner
function to show the nonclassicality in PAHS qualitatively and subsequently quantify
the negative volume of Wigner function as a measure of nonclassicality
\cite{kenfack2004negativity}.

The Wigner function of a pure quantum state in position-momentum space
is defined as \cite{wigner1932quantum}
\begin{equation}
\begin{array}{lcl}
W(x,p) & = & \frac{1}{\pi}\intop_{-\infty}^{\infty}\psi^{*}(x+y)\psi(x-y)\exp\left(2ipy\right)dy.\end{array}\label{eq:wigner-our}
\end{equation}
To compute Wigner function for PAHS we write the state in position
space using the definition of Fock state $|n\rangle$ in the position
space \cite{scully1999quantum}
\begin{equation}
|n\rangle=\phi_{n}(x)=b_{n}{\rm e}^{-\frac{x^{2}}{2}}H_{n}(x),\label{eq:fock-state in corrdinate space}
\end{equation}
where $b_{n}=\frac{1}{\sqrt{\pi^{\frac{1}{2}}2^{n}n!}}$ in the units
of $\hbar=1$, and $H_{n}(x)$ is the Hermite polynomial. Consequently,
we can express the finite dimensional qudit state (\ref{eq:qudit})
as 
\begin{equation}
|\psi_{d}(x)\rangle=N_{d}\sum_{n=0}^{d-1}c_{n}|n\rangle=\sum_{n=0}^{d-1}c_{n}\phi_{n}(x).\label{eq:qudit-position}
\end{equation}
 Using Eqs. (\ref{eq:wigner-our})-(\ref{eq:qudit-position}) we obtain
the Wigner function of PAHS as 
\begin{equation}
\begin{array}{lcl}
W(x,p) & = & \frac{{\rm e}^{-\left(x^{2}+p^{2}\right)}}{\pi}\sum\limits _{n,n^{\prime}=0}^{M}N_{{\rm PAHS}}^{2}\left[\tbinom{L\eta}{n}\tbinom{L\left(1-\eta\right)}{M-n}\tbinom{L\eta}{n^{\prime}}\tbinom{L\left(1-\eta\right)}{M-n^{\prime}}\right]^{\frac{1}{2}}\tbinom{L}{M}^{-1}\sqrt{\frac{\left(n+k\right)!\left(n^{\prime}+k\right)!}{n!n^{\prime}!}}\\
 & \times & (-1)^{n^{\prime}+k}\sqrt{2^{n^{\prime}-n}}\sqrt{\frac{(n+k)!}{(n^{\prime}+k)!}}(ip-x)^{n^{\prime}-n}L_{n+k}^{n^{\prime}-n}\left(2x^{2}+2p^{2}\right),\,\,\,\,\,n+k\leq n^{\prime}+k,
\end{array}\label{eq:Wig}
\end{equation}
where $L_{m}^{a}\left(z\right)$ is generalized Laguerre polynomial. 

\begin{figure}
\centering{}%
\includegraphics[scale=0.5]{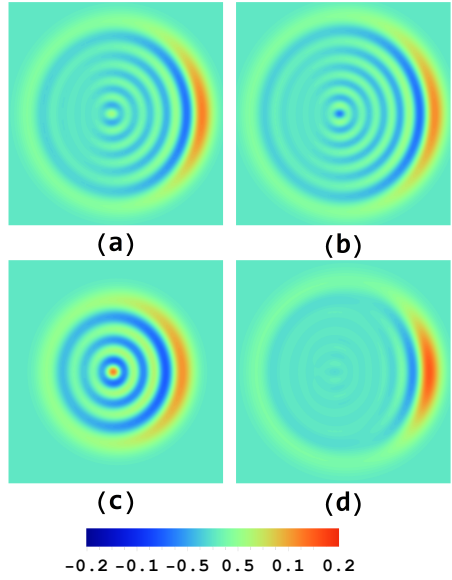}
\caption{\label{fig:wigner} (Color online)  {Wigner function of $k$-PAHS for
(a) $k=0$ with $M=10$, $\eta=0.9$, (b) $k=1$ with $M=10$, $\eta=0.9$,
(c) $k=1$ with $M=5$, $\eta=0.9$, and (d) $k=1$ with $M=10$,
$\eta=0.75$. All the contour plots are obtained for the same color scale (given at the bottom) with position $x$ and momentum $p$ variables shown in the horizontal and vertical axes, respectively.}}
\end{figure}

The Wigner function illustrates nonclassicality in PAHS by the negative
values in Fig. \ref{fig:wigner}. Specifically, we observed that the
negative values of the Wigner function are concentric circles, which
are also the signatures of non-Gaussian behavior of the Wigner function.
The number of rings of the negative Wigner function and non-Gaussian
features are decided by the various parameters of the state, i.e.,
the number of photon added, dimension, and probability $\eta$. With
an increasing dimension of the state the depth of Wigner minima decreases
while the number of rings increases (cf. Fig. \ref{fig:wigner} {(b)}
and (c)). Independently, photon addition increases the number of rings
(cf. Fig. \ref{fig:wigner} (a)-(b)), while Wigner function becomes
positive with a decrease in the value of $\eta$ (cf. Fig. \ref{fig:wigner}
{(b)} and (d)).

\color{black}

\section{Nonclassicality quantifiers and measures\label{sec:State-of-the}}

Here, we quantify the role of the non-Gaussianity inducing operation,
i.e., photon addition, in the enhancement of nonclassicality
in the finite dimensional intermediate state, i.e., PAHS. In what
follows, we report the {quantifiers} and measures of
nonclassicality with the relevant analytical expressions. To begin
with we will discuss two nonclassicality quntifiers (e.g., measure
to characterize quality of single photon source and anticlassicality)
and subsequently we will discuss nonclassicality measures (e.g., concurrence
potential and Wigner logarithmic negativity).

\begin{figure}
\begin{centering}
\begin{tabular}{cc}
\includegraphics[scale=0.5]{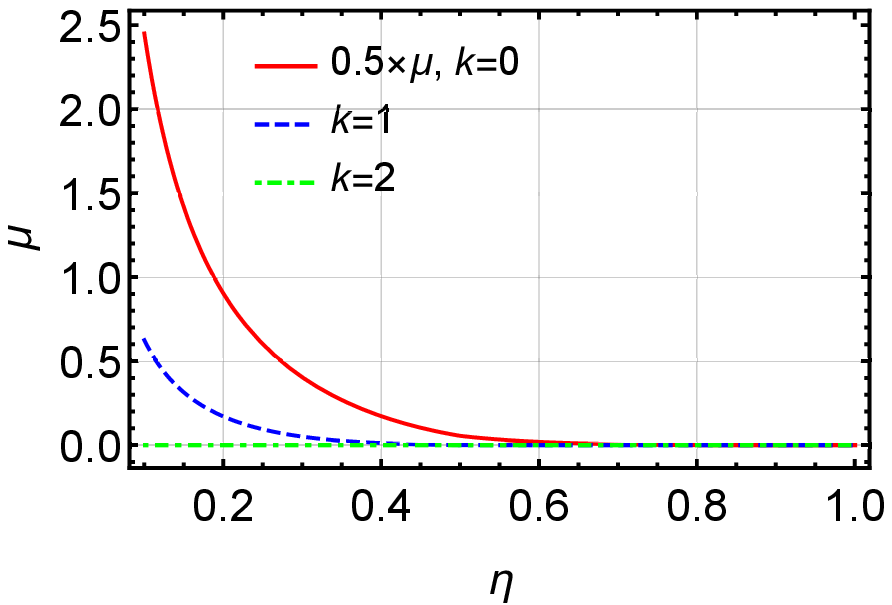} & \includegraphics[scale=0.5]{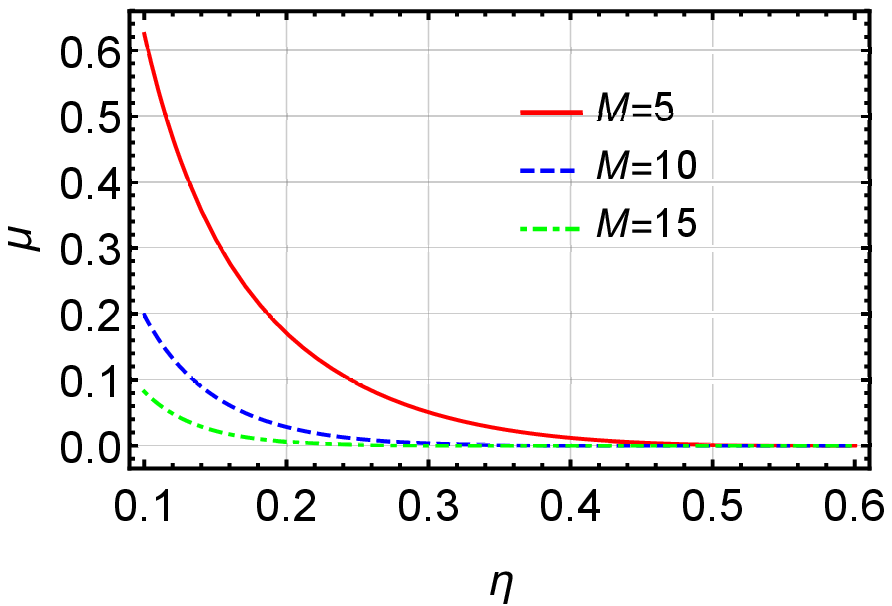} \tabularnewline
(a) & (b) \tabularnewline
\includegraphics[scale=0.5]{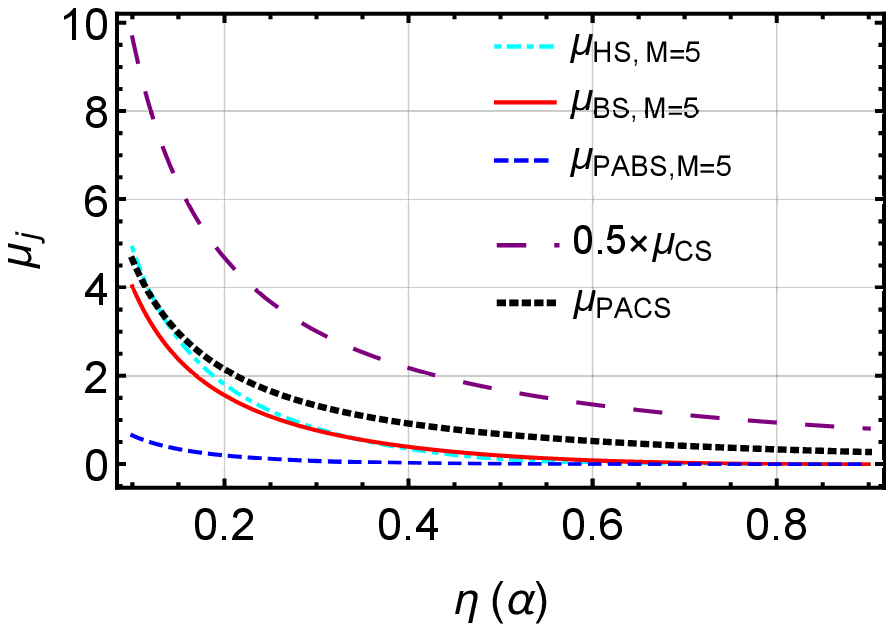}&  \includegraphics[scale=0.5]{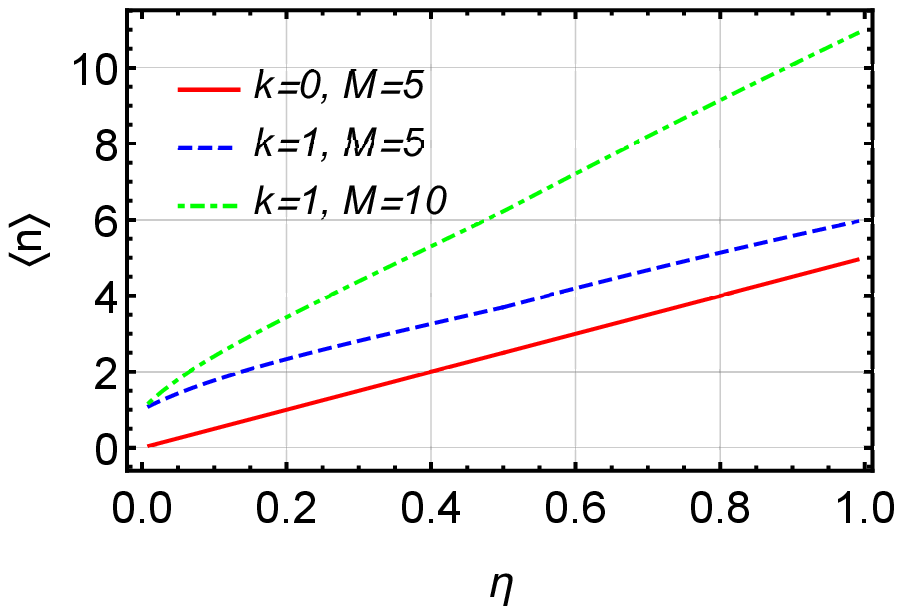}\tabularnewline
(c) & (d)\tabularnewline
\end{tabular}
\par\end{centering}
\caption{\label{fig:single-photon} (Color online) The dependence
of the measure to characterize the quality of single photon source
$\mu$ on (a) $\eta$ for different values of $k$ in PAHS with $M=5$,
(b) $\eta$ for different values of $M$ for single-PAHS. (c) $\mu$
is compared with that for photon added binomial state and binomial
state as function of $\eta$. In (c), we have also shown $\mu$ for
coherent state and single photon added coherent state with $\alpha=M\eta$
for $M\rightarrow\infty$. (d) Average
photon number $\left\langle n\right\rangle =\left\langle a^{\dagger}a\right\rangle $
with respect to $\eta$ for different $M$ and $k$ in PAHS.}
\end{figure}

\subsection{Measure to characterize quality of single photon source \label{subsec:Measure-to-characterize}}

\begin{figure}
\begin{centering}
\begin{tabular}{cc}
\includegraphics[scale=0.5]{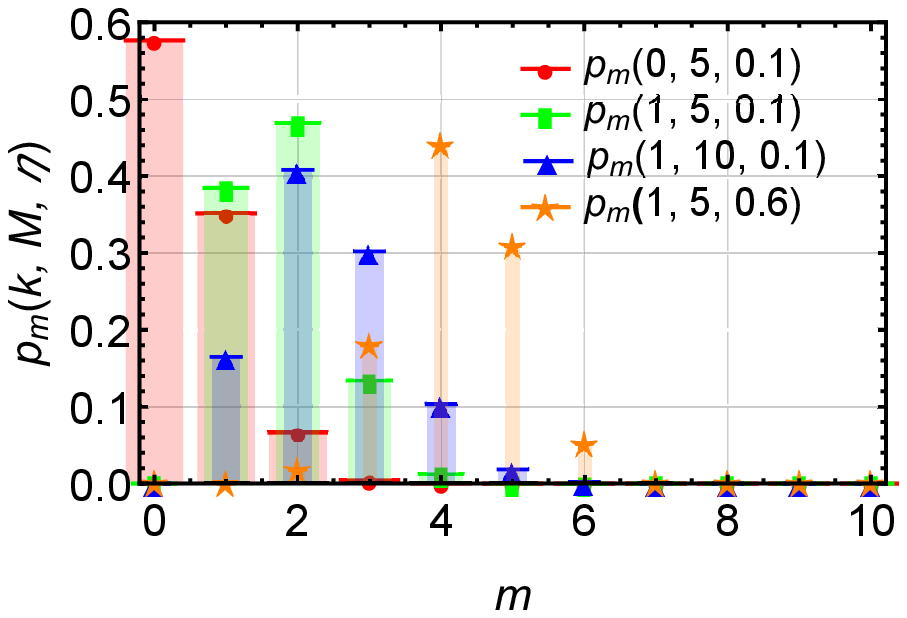} & \includegraphics[scale=0.5]{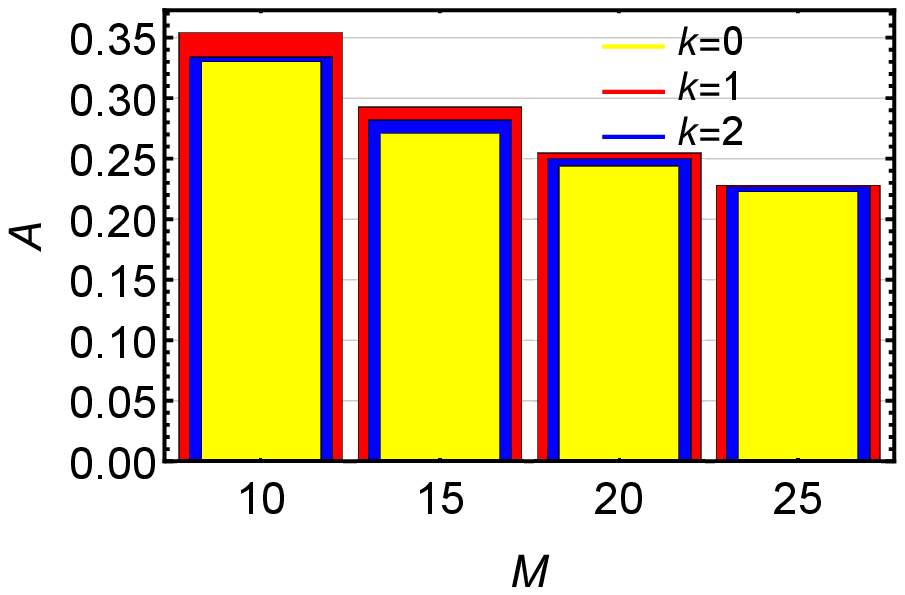} \tabularnewline
(a) & (b)\tabularnewline
\includegraphics[scale=0.5]{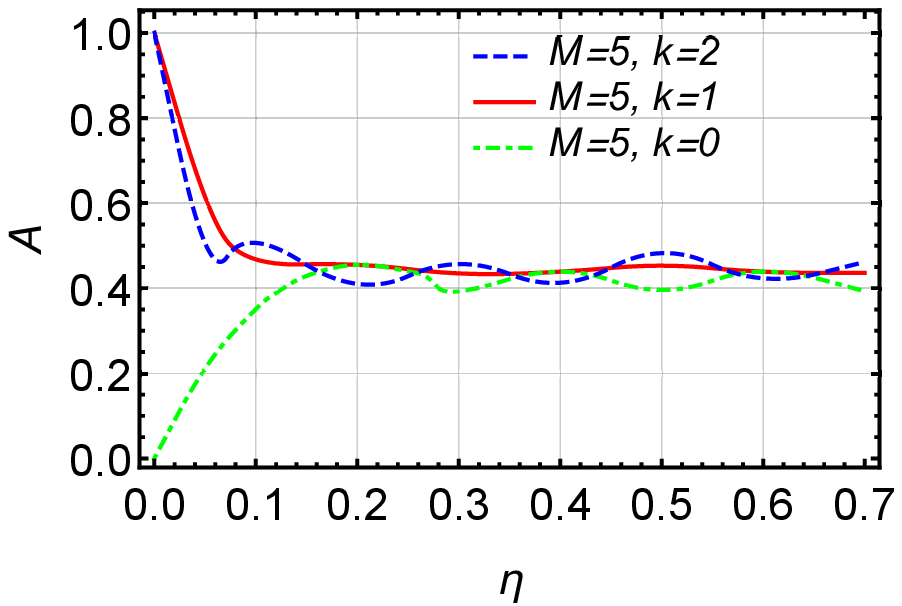} & \includegraphics[scale=0.5]{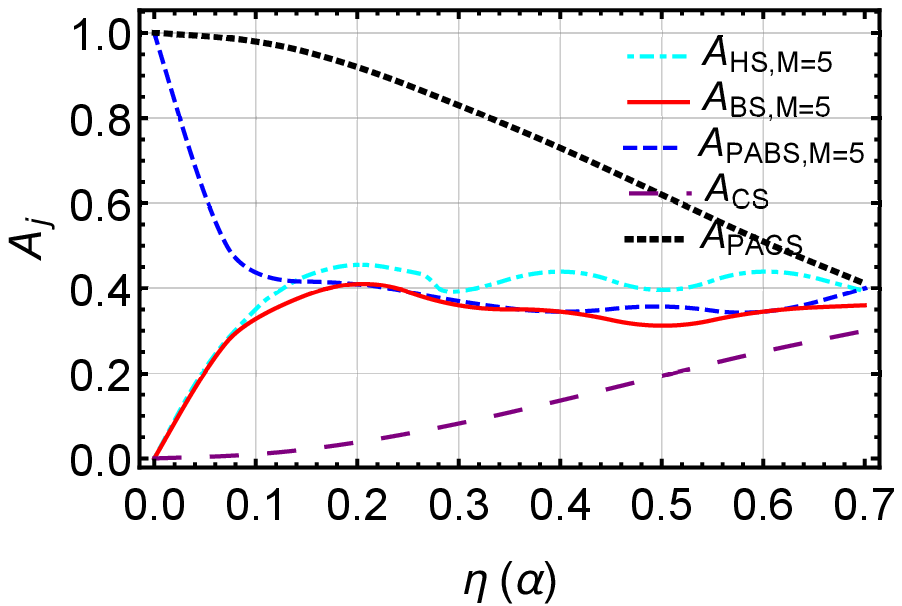}\tabularnewline
(c) & (d)\tabularnewline
\end{tabular}
\par\end{centering}
\caption{\label{fig:anticlassicality} (Color online) (a) The variation of 
photon number distribution $p_{m}$ with photon number $m$ for different
$\eta$, $M$, and photon addition $k$. Variation of
anticlassicality with (b) $M$ for different photon addition and
$\eta=0.18$ and (c) $\eta$.  {In (d), we have also shown $A$ for
binomial state and coherent state as well as their single photon added counterparts with $\alpha=M\eta$
for $M\rightarrow\infty$.}}
\end{figure}

A quantitative measure which can characterize the quality of the single
photon source \cite{pathak2010recent} can be defined mathematically
as

\begin{equation}
\mu=\frac{P_{1}}{1-\left(P_{0}+P_{1}\right)},\label{eq:Signle}
\end{equation}
where the probability $P_{i}$ of obtaining exactly $i$ photons in
a pulse is obtained from the corresponding photon number distribution
$p_{m}$ as $P_{i}=p_{m}\left(m=i\right)\forall i\in\left\{ 0,1\right\} $.
Specifically, this quantifies the ratio of single photon pulses with
multiphoton pulses in the concerned state and is expected to be large
for a desirable quantum state of light for single photon generation.
The more is the value of $\mu$ for a state, the better it is as a
single photon source to be used in quantum cryptography. In case of
PAHS, the photon number distribution $p_{m}^{{\rm PAHS}}$ can be
obtained as 
\begin{equation}
\begin{array}{lcl}
p_{m}^{{\rm PAHS}} & = & \left|\left\langle m\mid L,M,\eta,k\right\rangle \right|^{2}\\
 & = & \begin{cases}
\left|N_{{\rm PAHS}}\tbinom{L\eta}{m-k}^{\frac{1}{2}}\tbinom{L\left(1-\eta\right)}{M-m+k}^{\frac{1}{2}}\tbinom{L}{M}^{-\frac{1}{2}}\sqrt{\frac{m!}{(m-k)!}}\right|^{2}, & m\geq k\\
0, & m< k.
\end{cases}
\end{array}\label{eq:SPS-PAHS}
\end{equation}
We have already mentioned that results corresponding to a set of quantum
states can be obtained from Eq. (\ref{eq:SPS-PAHS}) in the limiting
cases. Notice that the probability of detecting a pulse having photon
number less than the number of photon added is zero in Eq. (\ref{eq:SPS-PAHS}).
Thus, photon addition ($k>1$) is certainly not providing any advantage
in the context of single photon generation (cf. Fig. \ref{fig:single-photon}
(a)). As far as $k=1$ is concerned, it also reduces the probability
of single photon pulses with respect to the corresponding
hypergeometric state {(except at $\eta\rightarrow0$ and $L\rightarrow\infty$ when the photon added state reduces to Fock state $\left|n=1\right\rangle$)}. This is physically expected and is consistent
with our intuition. Additionally, with increase in the dimension of
qudit, average photon number $\left\langle n\right\rangle =\left\langle a^{\dagger}a\right\rangle $
increases (shown in Fig. \ref{fig:single-photon} (d)) because
the photon number distribution has altered. In other words, the probability for higher
photon number pulses is increasing at the cost of that for the lower photon
number pulses (shown in Fig. \ref{fig:anticlassicality} (a)), and
therefore, the probability of single photon pulses decreases (cf.
Fig. \ref{fig:single-photon} (b)). We also observed that hypergeometric
state performs better than binomial state as a
single photon source for small values of $\eta$ (cf. Fig. \ref{fig:single-photon} (c)). Further
photon addition in all these cases just reduces the quality of these
states to be used as a single photon source. This can also be visualized
from the photon number distribution of hypergeometric state and PAHS
in Fig. \ref{fig:anticlassicality} (a) that due to a single photon
addition probability of zero photon pulses becomes zero. We have not
discussed the case of $k>1$ as the probability of both zero and single
photon pulses becomes zero in that case. For the
sake of completeness we have also shown dependence of the average photon number on all the state parameters in Fig. \ref{fig:single-photon} (d) which illustrates a monotonous increase in $\left\langle n\right\rangle $.

{ It is worth mentioning that the findings for an infinite dimensional state
(coherent and photon added coherent states in this case) cannot be compared trivially with
that of a finite dimensional state. However, in the present case, photon
addition reduces the quality of single photon source from the coherent state as observed in case of PAHS. In view of the relatively difficult state preparation, one may not be too optimistic to use the present state as approximate single photon source.  Notice that a single photon added states reduces to ideal single photon source when the state parameter $\eta$ or $\alpha$ is close to zero. Thus, these photon added states should be a better candidate for single photon source in comparison to the corresponding parent state but this quantifier fails to capture this behavior in Fig. \ref{fig:single-photon} (c).} {However, the intrinsic beauty of this
relatively simple measure will be further illustrated in the next
subsection where we will explore a connection between $\mu$ and
anticlassicality, a well studied concept introduced by Dodonov et
al. in 2003 \cite{dodonov2003classicality}. The connection may add
to the existing physical meaning of anticlassicality.}

\subsection{Anticlassicality \label{subsec:Anticlassicality}}

Dodonov et al., introduced a measure of nonclassicality as the distance
between the nonclassical state to be studied and Fock state (the most
nonclassical state) \cite{dodonov2003classicality}. Mathematically,
anticlassicality for an arbitrary quantum state $\rho$ is defined
as 

\begin{equation}
A=\underset{m>0}{{\rm max}}\ p_{m}^{\rho}\label{anticlassicality-def}
\end{equation}
in terms of its photon number distribution $p_{m}^{\rho}$. In our
case, $\rho$ is PAHS and therefore we use $p_{m}^{{\rm PAHS}}$ as
defined in Eq. (\ref{eq:SPS-PAHS}) to obtain anticlassicality using
Eq. (\ref{anticlassicality-def}). {Incidentally, the variation of
anticlassicality measure shows that increasing $k$, $M$, and $\eta$
have the same effect: anticlassicality is found to decrease with increase
in all these parameters (cf. Fig. \ref{fig:anticlassicality} (b)
and (c)). This is consistent with the observations in \cite{dodonov2003classicality}
that anticlassicality decreases with the increasing average photon
number $\left\langle n\right\rangle =\left\langle a^{\dagger}a\right\rangle >1$
while it increases for $0<\left\langle n\right\rangle <1$.} {However, for the large values of all these parameters anticlassicality does not change considerably and thus becomes comparable as observed in Fig. \ref{fig:anticlassicality} (b)
and (c).} 
{  Interestingly, for $\eta=0$ PAHS behaves like Fock $\left|n=1\right\rangle$ and vacuum for $k=1$ and 0, respectively. Thus, anticlassicality is unity (zero) in the former (latter) case. In Ref.~\cite{dodonov2003classicality}, it is shown that allowing $m=0$ as well in the definition (\ref{anticlassicality-def}) the parameter will be unity even in the latter case. In this scenario, we will observe that modified parameter for both HS and PAHS to decrease with $\eta$ as well similar to that observed in variation of $\mu$. Therefore, the observed behavior of anticlassicality maximized over complete Fock basis is similar to that of the measure of quality of single photon source. 
We also observed that anticlassicality of hypergeometric
state is more (less) than that of the binomial state for small (large) values of $\eta$ (cf. Fig. \ref{fig:anticlassicality} (d)). Further
photon addition in all these cases increases the average photon number and anticlassicality can be observed to decrease with $\eta$.}

\begin{figure}
\centering{}%
\begin{tabular}{cc}
\includegraphics[scale=0.5]{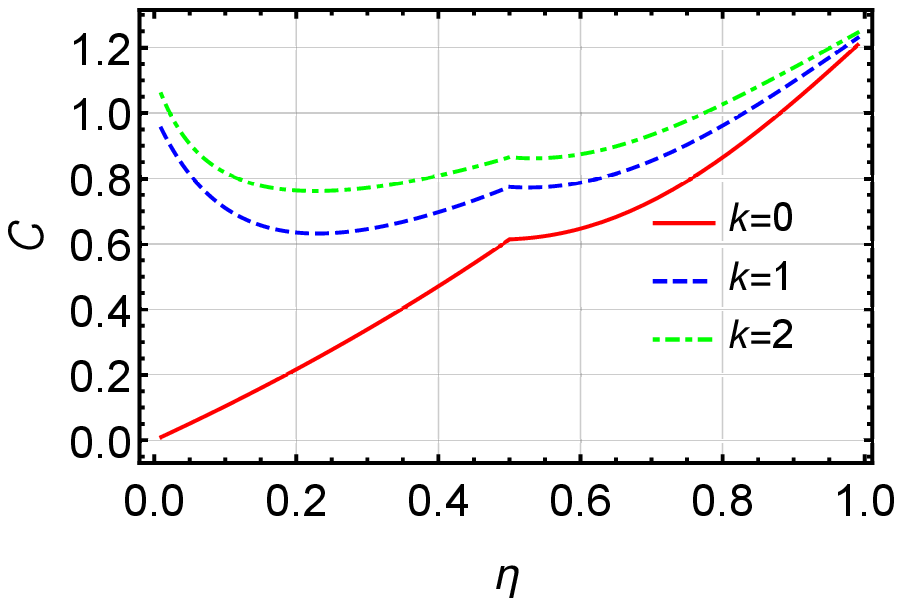} & \includegraphics[scale=0.5]{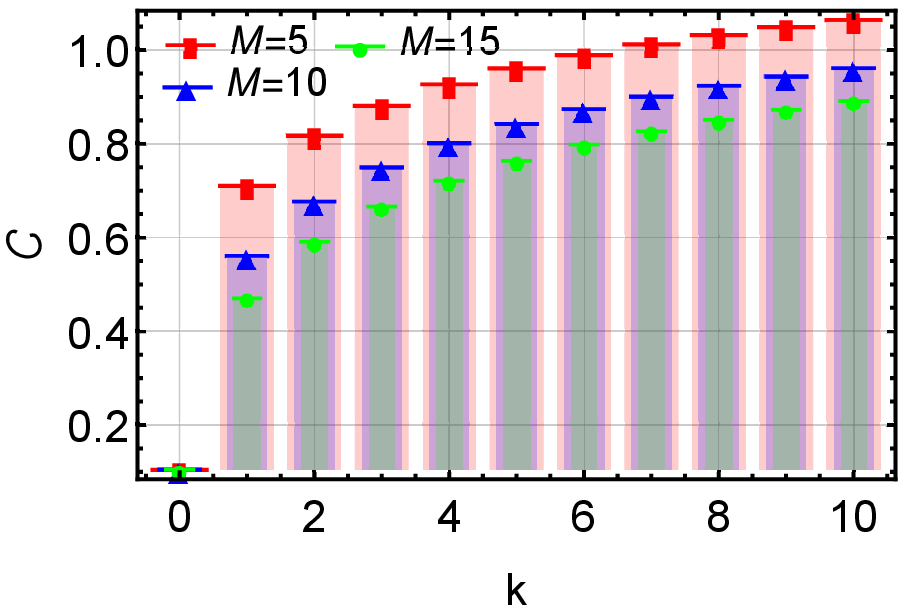}\tabularnewline
(a) & (b)\tabularnewline
\end{tabular}\caption{\label{fig:Concurrence}(Color online) Concurrence potential for $k$-PAHS
as a function of (a) $\eta$ with $M=5$ and (b) photon addition with $\eta=0.1$. }
\end{figure}

\subsection{Concurrence potential \label{subsec:Concurrence}}

There are several measures of nonclassicality proposed for the purpose
of quantitative analysis of nonclassicality in a quantum state. All
these measures have some inherent limitations (see \cite{miranowicz2015statistical}
for discussion). These measures are often not monotone of each other
and therefore, it becomes important to study either more than one
nonclassicality measure or mention the quantifier used. One such measure,
that we are going to use here, attempted to quantify the amount of
nonclassicality present in an arbitrary quantum state using entanglement
measures \cite{asboth2005computable}, namely entanglement potential.
However, entanglement measures are also not monotone of each other.

The idea of entanglement potential \cite{asboth2005computable,vogel2014unified,ge2015conservation,miranowicz2015increasing,arkhipov2016interplay} is based on the fact that classical
states cannot generate entanglement in the output of the beamsplitter,
while mixing of an arbitrary nonclassical state with any classical
state (for example vacuum) on a beamsplitter generates an equivalent
amount of entanglement in the output. Without loss of generality,
for the total input state of the beamsplitter $\left(|\psi\rangle\otimes|0\rangle\right)=\left(\stackrel[n=0]{M}{\sum}c_{n}|n\rangle\right)\otimes|0\rangle$,
the output of the beamsplitter can be expressed in the Fock basis
as 
\begin{equation}
|\Phi\rangle=U\left(|\psi\rangle\otimes|0\rangle\right)\equiv U|\psi,0\rangle=\sum_{n=0}^{M}\,\frac{c_{n}}{2^{n/2}}\sum_{j=0}^{n}\sqrt{\tbinom{n}{j}}\,\,|j,\,n-j\rangle.\label{eq:inout_psi}
\end{equation}
Similarly, mixing PAHS and vacuum state at the beamsplitter, the post
beamsplitter state can be written as 

\begin{equation}
|\phi\rangle=N_{{\rm PAHS}}\sum\limits _{n=0}^{M}\left[\tbinom{L\eta}{n}\tbinom{L\left(1-\eta\right)}{M-n}\right]^{\frac{1}{2}}\tbinom{L}{M}^{-\frac{1}{2}}\frac{1}{\sqrt{n!}}\sum\limits _{k_{1}=0}^{n+k}\tbinom{n+k}{k_{1}}\left(\frac{1}{\sqrt{2}}\right)^{k_{1}}\left(\frac{i}{\sqrt{2}}\right)^{n+k-k_{1}}\sqrt{k_{1}!\left(n+k-k_{1}\right)!}|k_{1},\,n+k-k_{1}\rangle,\label{postBSstate}
\end{equation}
where $i^{2}=-1$.
Concurrence is formally defined for a bipartite state $\rho_{AB}$
as \cite{asboth2005computable}

\begin{equation}
C=\sqrt{2\left(1-{\rm Tr}\left(\rho_{B}^{2}\right)\right)},\label{concurrence}
\end{equation}
where $\rho_{B}$ is obtained after tracing over subsystem $B$. Mathematical
expression for ${\rm Tr}\left(\rho_{B}^{2}\right)$, in the present
case with $\rho_{AB}=|\phi\rangle\langle\phi|$, is 

\begin{equation}
\begin{array}{lcl}
{\rm Tr}\left(\rho_{B}^{2}\right) & = & N_{{\rm PAHS}}^{4}\sum\limits _{n=0}^{M}\sum\limits _{m=0}^{M}\sum\limits _{r=0}^{M}\left[\tbinom{L\eta}{n}\tbinom{L\left(1-\eta\right)}{M-n}\tbinom{L\eta}{m}\tbinom{L\left(1-\eta\right)}{M-m}\tbinom{L\eta}{r}\tbinom{L\left(1-\eta\right)}{M-r}\tbinom{L\eta}{n-m+r}\tbinom{L\left(1-\eta\right)}{M-n+m-r}\right]^{\frac{1}{2}}\\
 & \times & \left(\frac{1}{2}\right)^{n+2k+r}\tbinom{L}{M}^{-2}\sum\limits _{k_{1}=0}^{n+k}\frac{\left(n+k\right)!\left(m+k\right)!\left(r+k\right)!\left(n-m+r+k\right)!}{\sqrt{n!m!r!\left(n-m+r\right)!}k_{1}!\left(n+k-k_{1}\right)!\left(k+m-k_{1}\right)!\left(r-m+k_{1}\right)!}.
\end{array}\label{concurrence-analytical}
\end{equation}
This can be used to obtain an analytic expression for concurrence
potential. The value of concurrence potential equal to $0$ (nonzero)
corresponds to a separable (entangled) state, i.e., the input state
is classical (nonclassical). 

{  In contrast to anticlassicality and the measure for the characterization of single photon
source, in this case, we can clearly see the advantage of increasing photon addition
and $\eta$ in enhancing the amount of
nonclassicality present in the state. Specifically, nonclassicality can be observed to increase here with these state parameters by the corresponding large values of concurrence
potential (see Fig. \ref{fig:Concurrence}).} Notice that for higher
values of photon addition nonclassicality decreases with increasing
probability $\eta$ when $\eta\approx0$ because the quantum state
in this case is very close to Fock state. However, the decrease in
the dimension of PAHS keeping the rest of the parameters unchanged
also increases nonclassicality of the state due to holeburning in
the state, which is known as nonclassicality inducing/enhancing operation. There is a change in the slope of the curve of concurrence potential
with $\eta$ for $\eta=0.5$ as the derivative of $C$ is not defined
because of parameter $L$. This feature is often studied as a sudden
change of entanglement with significance in quantum information processing
\cite{celeri2010sudden,siyouri2016negativity}. 

\subsection{Wigner logarithmic negativity \label{subsec:Wigner-log-neg}}

A measure of nonclassicality was introduced which quantifies the negative
volume of Wigner function \cite{kenfack2004negativity}. More recently
a resource theoretic counterpart of Wigner volume has been proposed
as Wigner logarithmic negativity \cite{albarelli2018resource}. Here,
we aim to quantify the amount of nonclassicality illustrated by the
Wigner function (\ref{eq:Wig}) using Wigner logarithmic negativity
\cite{albarelli2018resource}, which is defined as
\begin{equation}
W\left(\delta\right)=\log\left(\int\int dxdp\left|W\left(x,p\right)\right|\right),\label{eq:WLN}
\end{equation}
where the integration is performed over both position and momentum
quadratures in the phase space. 

\begin{figure}
\centering{}%
\begin{tabular}{ccc}
\includegraphics[scale=0.4]{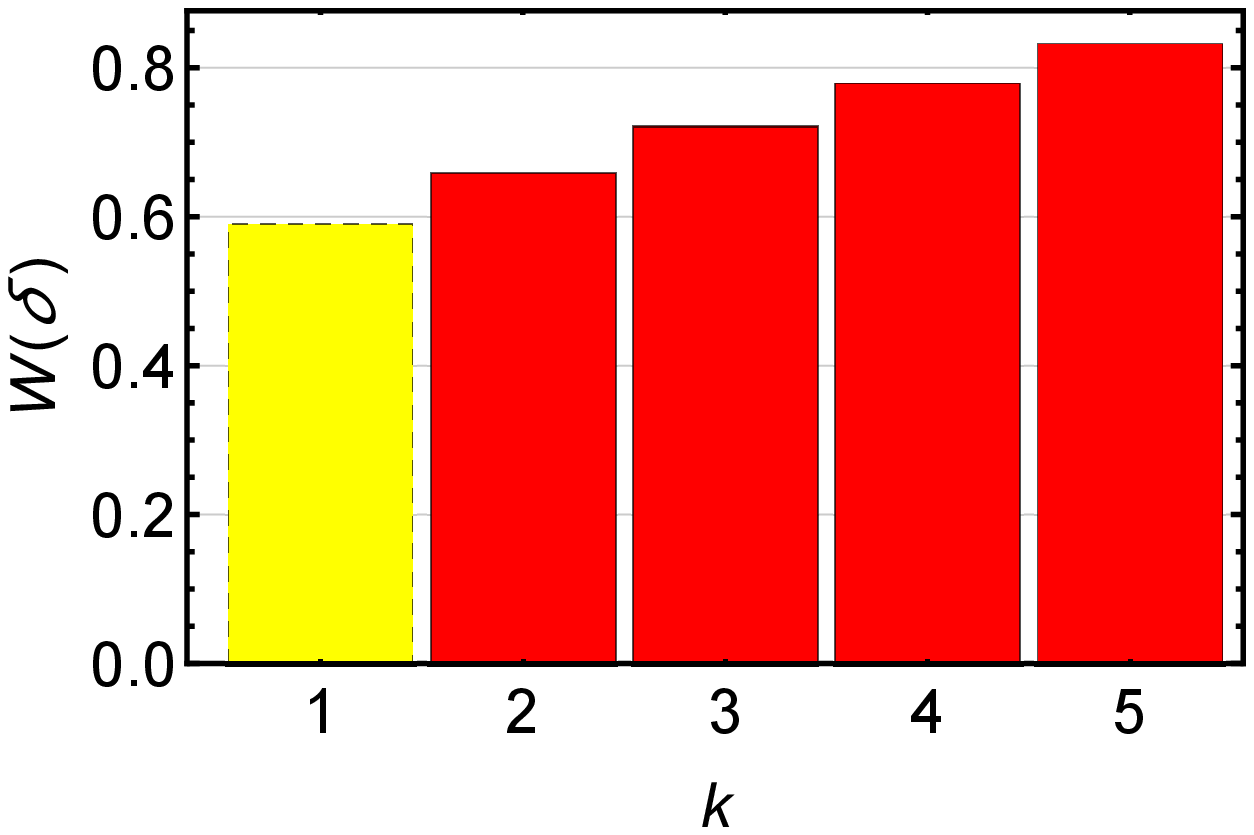} & \includegraphics[scale=0.57]{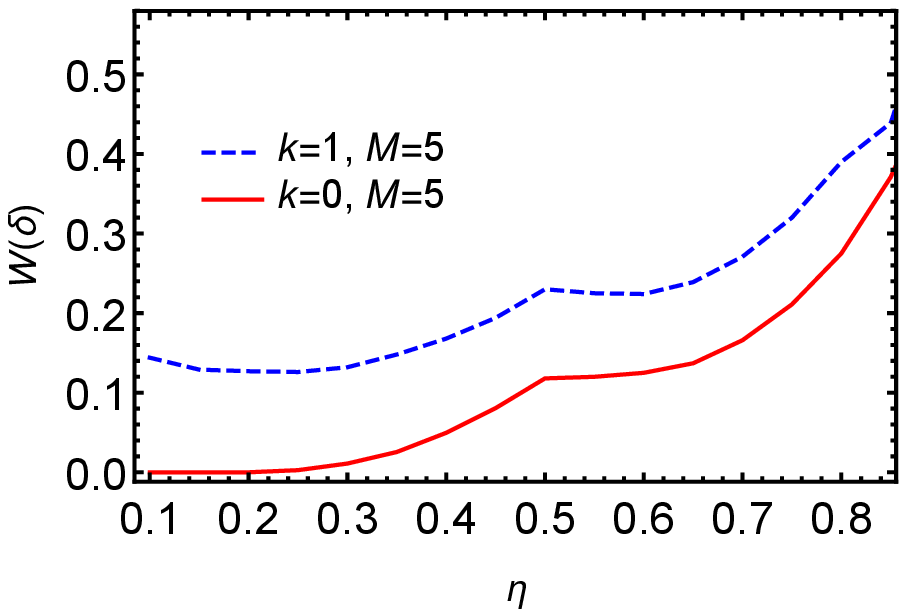} & \includegraphics[scale=0.4]{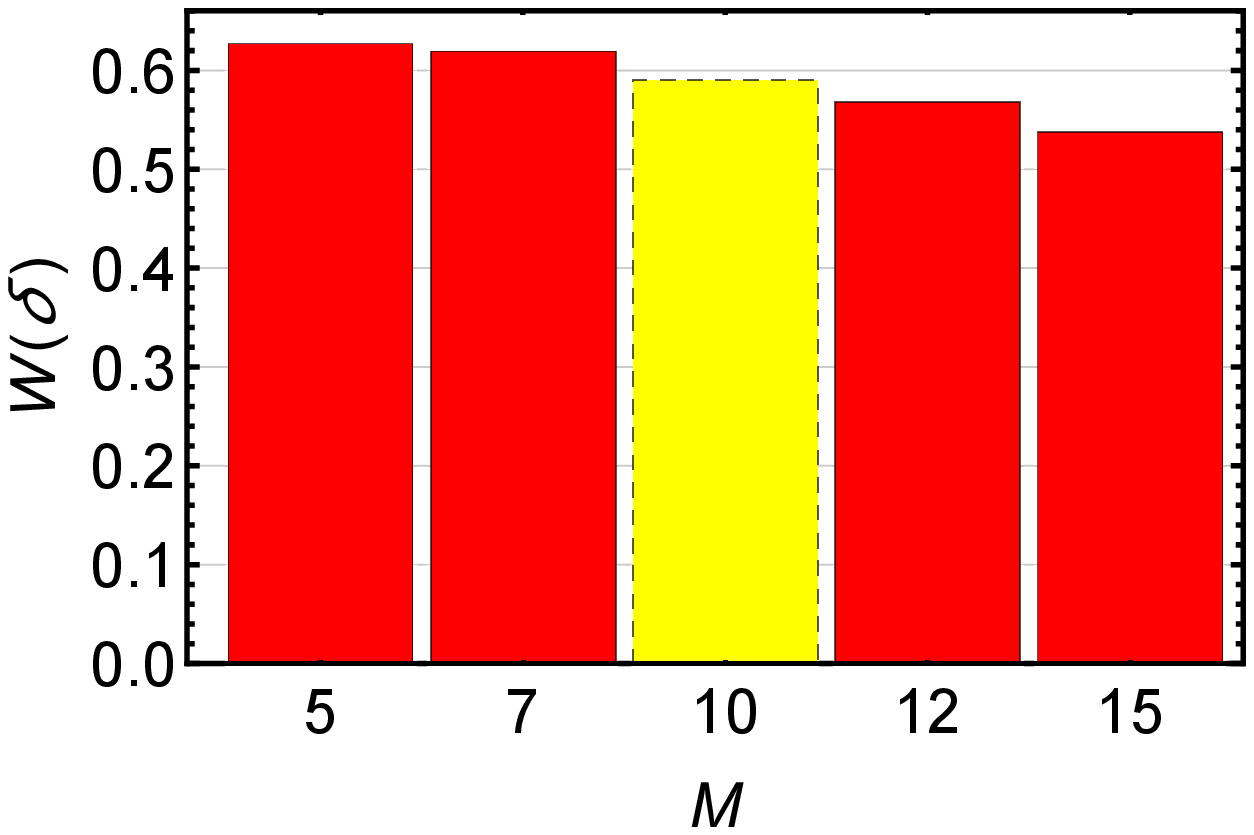}\tabularnewline
(a) & (b) & (c)\tabularnewline
\end{tabular}\caption{\label{fig:wig-log-neg}(Color online) The variation of Wigner logarithmic
negativity as a function of (a) photon addition for $M=10$ and $\eta=0.9$,
(b) $\eta$, and (c) $M$ for $k=1$ and $\eta=0.9.$  {The yellow colored bar (with dashed lines) in (a) and (c) are obtained for the same set of values of parameters.}}
\end{figure}

All the qualitative observations regarding the dependence of Wigner
negativity on $k$, $M$, and $\eta$ are established quantitatively
using Wigner logarithmic negativity. Specifically, the Wigner logarithmic
negativity increases with photon addition and $\eta$, while increasing
dimension reduces the nonclassicality (see Fig. \ref{fig:wig-log-neg}). Specifically, the yellow colored bars (with dashed lines) in Fig. \ref{fig:wig-log-neg} (a) and (c) is obtained for single-PAHS of dimension with $M=10$, which can be observed to increase with increase in photon addition and/or decrease in the dimension of PAHS.
Interestingly, the concurrence potential of the two-mode quantum state
generated from PAHS at a beamspitter (in Fig. \ref{fig:Concurrence})
and the Wigner logarithmic negativity (in Fig. \ref{fig:wig-log-neg})
gave us the same conclusion. 

Thus, both measures of nonclassicality used here show that nonclassicality
can be increased by increasing value of $\eta$ and/or adding single
photons in the state which creates hole in the photon number distribution
from lower side and/or reducing dimension of the state, i.e., burning
holes for the large $n$ states.

\section{Results and concluding remarks \label{sec:Result-and-concluding}}

We proposed PAHS state with an aim to quantify the amount of nonclassicality
in the concerned state and establish the role of quantum state engineering
tools, namely, photon addition and holeburning, in enhancement of
nonclassicality in hypergeometric state. Here, we used a measure to
characterize the quality of single photon sources 
and anticlassicality, both of which showed us that these nonclassical
features depend on the average photon numbers. Specifically, with
the photon addition, dimension, and probability $\eta$ the average
photon number increases as the photon number distribution at higher
photon numbers increase at the cost of that for lower photon numbers.
Therefore, in view of the present results, we can conclude that anticlassicality
is related to a measure for the characterization
of single photon source. Interestingly, photon subtraction is also
known to increase the average photon number of a quantum state. Therefore,
it would be worth analyzing the role of photon subtraction in changing
anticlassicality of a quantum state, especially when a hole in the
photon number distribution at vacuum is not expected due to single
photon subtraction unlike photon addition.

We have further quantified nonclassicality in PAHS
using Wigner logarithmic negativity and concurrence potential, i.e.,
the entanglement generated at the beamspitter. Specifically, we observed
that nonclassicality increases by non-Gaussianity inducing operation
photon addition and increasing probability parameter $\eta$, while
nonclassicality reduces by increasing the dimension of the state.
In the attempt of obtaining Wigner logarithmic negativity we also
characterized nonclassicality qualitatively using the Wigner function
which illustrated the non-Gaussian behavior of the quantum state under
study.

The present results illustrate the highly nonclassical behavior of
the PAHS, which can be enhanced by photon addition, burning some holes
in the phase space (i.e., reducing the dimension of the state), and
optimizing the value of parameter $\eta$. The choice of the present
state allows us to reduce the corresponding results for a set of quantum
states, for instance, binomial state, phase independent coherent state,
and respective photon added states as well as hypergeometric state.
This study can be further extended to the role of the 
rest of the quantum state engineering tools in enhancing nonclassicality
and non-Gaussianity of the other finite dimensional quantum states.
We hope the present work will motivate such study and find applications
in some quantum information processing tasks, specially in quantum
cryptography where at one hand single photon sources (which can be
generated using states obtained as limiting cases of PAHS) are required
for discrete variable quantum key distribution and on the other hand
continuous variable quantum cryptography useful for metropolitan quantum
network can be realized using these states.

\section*{Acknowledgment}

PM and AP acknowledge the support from DRDO, India project no.\textcolor{black}{{}
ANURAG/MMG/CARS/2018-19/071.} KT acknowledges the financial support
from the Operational Programme Research, Development and Education
- European Regional Development Fund project no. CZ.02.1.01/0.0/0.0/16\_019/0000754
of the Ministry of Education, Youth and Sports of the Czech Republic.

\end{document}